\documentclass{ieeeaccess}
\usepackage{cite}
\usepackage{amsmath,amssymb,amsfonts}
\usepackage{algorithmic}
\usepackage{graphicx}
\usepackage{textcomp}
\usepackage{xspace}
\usepackage{booktabs} 
\usepackage{comment}
\usepackage{makecell}
\usepackage{url}
\usepackage{subfigure}
\usepackage[utf8]{inputenc}
\usepackage{newunicodechar}
\usepackage{verbatim}
\usepackage[utf8]{inputenc}
\usepackage{pifont}
\usepackage[dvipsnames]{xcolor}

\usepackage{bm}
\makeatletter
\AtBeginDocument{\DeclareMathVersion{bold}
\SetSymbolFont{operators}{bold}{T1}{times}{b}{n}
\SetSymbolFont{NewLetters}{bold}{T1}{times}{b}{it}
\SetMathAlphabet{\mathrm}{bold}{T1}{times}{b}{n}
\SetMathAlphabet{\mathit}{bold}{T1}{times}{b}{it}
\SetMathAlphabet{\mathbf}{bold}{T1}{times}{b}{n}
\SetMathAlphabet{\mathtt}{bold}{OT1}{pcr}{b}{n}
\SetSymbolFont{symbols}{bold}{OMS}{cmsy}{b}{n}
\renewcommand\boldmath{\@nomath\boldmath\mathversion{bold}}}
\makeatother

\newcommand\approach{{\sc PR4Code}\xspace}

\def\BibTeX{{\rm B\kern-.05em{\sc i\kern-.025em b}\kern-.08em
    T\kern-.1667em\lower.7ex\hbox{E}\kern-.125emX}}

\begin{document}
\history{Date of publication xxxx 00, 0000, date of current version xxxx 00, 0000.}
\doi{10.1109/ACCESS.2024.0429000}

\title{\approach: A Pull Requests Dataset for AI Code Generation}
\author{\uppercase{Benedetta Donato}\authorrefmark{1} \IEEEmembership{Student Member, IEEE},
\uppercase{Leonardo Mariani}\authorrefmark{2} \IEEEmembership{Senior Member, IEEE},
\uppercase{Daniela Micucci}\authorrefmark{3} \IEEEmembership{Member, IEEE}, and 
\uppercase{Oliviero Riganelli}\authorrefmark{4}}

\address[1]{University of Milano-Bicocca (e-mail: benedetta.donato@unimib.it)}
\address[2]{University of Milano-Bicocca (e-mail: leonardo.mariani@unimib.it)}
\address[3]{University of Milano-Bicocca (e-mail: daniela.micucci@unimib.it)}
\address[4]{University of Milano-Bicocca (e-mail: oliviero.riganelli@unimib.it)}

\markboth
{Donato \headeretal: \approach: A Pull Requests Dataset for AI Code Generation}
{Donato \headeretal: \approach: A Pull Requests Dataset for AI Code Generation}

\corresp{Corresponding author: Benedetta Donato (e-mail: benedetta.donato@unimib.it).}

\begin{abstract}
AI‑driven code generation is evolving from handling small, localized code‑completion tasks toward supporting the implementation of full software features. To support this transition, this paper introduces  {\approach}, a dataset of 
\textbf{4,508 Java} and \textbf{8,831 Python} curated Pull Requests (PRs) collected from GitHub, each enriched with metadata, commit histories, and detailed code changes.
Unlike resources focused on isolated code fragments, \approach captures the feature implementation process, delivering a large and diverse collection of real-world development tasks that can be used to assess AI-driven solutions.
Our analysis reveals substantial variability in PR structure, commit granularity, and textual content. In addition, we release scripts to regenerate and update the dataset, ensuring reproducibility and maintainability.
\end{abstract}


\titlepgskip=-21pt

\maketitle

\section{Introduction}\label{sec:introduction}
\PARstart{S}{oftware} evolution is a continuous process that results in a stream of code changes produced by developers. Each modification reflects a decision that embodies how a problem was understood and solved, for instance, to add, modify, or improve functionalities. In recent years, AI-based systems have been increasingly adopted to support and accelerate a wide range of software 
development activities.

Recent advances in Large Language Models (LLMs), such as GPT-5~\cite{openai2025gpt5}, Claude~\cite{anthropic2024claude}, and specialized code-oriented models such as CodeLlama~\cite{roziere2023code} and CodeGemma~\cite{codegemma2024}, have demonstrated remarkable capabilities in generating code from natural-language specifications. These models are now embedded in widely adopted industrial tools, including GitHub Copilot~\cite{copilot2024}, Amazon CodeWhisperer~\cite{aws2024amazonqdeveloper}, and Tabnine~\cite{tabnine2024}. Empirical studies report measurable productivity gains associated with the adoption of such tools in real-world development settings~\cite{davila2024industry,peng2023impact,cui2025effects}.

Despite these advances, current AI-based approaches remain largely effective only for relatively small-scale tasks, such as code completion~\cite{izadi2024language,Husein2025LLMCodeCompletion,Wu2025RepoMasterEval} and method-level generation~\cite{corso2024generating,Fakhoury2024TiCoder,fagadau2024analyzing}. Their performance drops significantly when the goal is to generate larger and more complex code units, including full features, modules, or applications. Such tasks require a deeper understanding of software architecture, cross-file dependencies, and integration with existing components, aspects that current models still struggle to handle effectively~\cite{jimenez2023swe,corso2024generating,Sagodi2024RealityCheck,Dong2025LLMAgentsSurvey}.

A major obstacle to systematically studying these limitations is the lack of datasets that capture feature-level software development in realistic settings. Existing benchmarks and corpora often focus on isolated code changes~\cite{10.1145/2610384.2628055}, single commits~\cite{monperrus2021megadiff}, or narrowly scoped bug-fixing tasks~\cite{Karampatsis_2020}, %
and therefore fail to represent the broader development context in which real-world changes 
are conceived, iteratively refined, and integrated across multiple artifacts. In fact, feature development is typically documented through natural-language descriptions that may be incomplete or cross-referenced, implemented via multiple commits, and realized through coordinated changes spanning source code, tests, configuration files, and documentation.

To address this gap, this paper introduces \emph{PR4Code}, a large-scale dataset of real-world pull requests collected from popular Java and Python GitHub repositories. PR4Code adopts pull requests as the primary unit of analysis, as they naturally encapsulate the lifecycle of a development task, from its textual description to the sequence of commits and the final integrated changes. For each pull request, the dataset provides rich contextual information, including natural-language descriptions at both PR and commit levels, fine-grained and cumulative code diffs, and metadata that make it possible to reconstruct the development history and the surrounding context.

By construction, PR4Code enables the investigation of fundamental research questions concerning AI-based code generation in realistic software engineering scenarios. In particular, the dataset makes it possible to study whether current AI-based approaches can autonomously implement non-trivial software features within existing projects, how effectively they can handle development tasks involving heterogeneous artifacts beyond source code (e.g., tests and configuration files), and to what extent cross-referenced or non-self-contained natural-language descriptions affect the quality and correctness of generated code changes. These questions are difficult to address using existing benchmarks, yet are central to understanding the practical limitations of LLM-based systems in feature-level development.

The contributions of this paper are threefold. First, we present PR4Code, a curated dataset comprising 13,339 pull requests and 46,334 associated commits that span 1,055 Java and Python repositories. Second, we provide a detailed characterization of the dataset, analyzing both textual descriptions and code changes to highlight the diversity and complexity of real-world feature development tasks. Third, we release the scripts used to construct the dataset, enabling reproducibility and facilitating future extensions aligned with the evolving landscape of AI-based software development tools.

The paper is organized as follows. Section~\ref{sec:related} discusses our work in the context of related studies. Section~\ref{sec:methodology} describes the methodology we used to build the \approach dataset. Section~\ref{sec:approach} presents the \approach dataset. Section~\ref{sec:discussion} discusses some exemplary uses of the dataset. Section~\ref{sec:usage} reports early results from a small-scale experiment. Finally, Section~\ref{sec:conclusions} provides final remarks.

\section{Related Work} \label{sec:related}
The problem of evaluating LLMs and AI-based systems on realistic software engineering tasks has motivated the creation of a variety of datasets and benchmarks, each targeting different granularities of code change and development contexts. 

Early resources mainly focused on isolated code fragments or single-function modifications. HumanEval{~\cite{chen2021evaluating}} evaluates the ability of models to generate correct function implementations from short natural-language prompts using test-based metrics such as pass@k. Recent extensions such as HumanEval-XL{~\cite{peng2024humaneval}} broaden this setting to multilingual scenarios while still preserving a function-level generation paradigm. More recently, benchmarks such as HumanEval Pro and MBPP Pro{~\cite{yu2024humaneval}} introduced self-invoking code generation tasks, requiring models to reuse previously generated code to solve more complex problems. These works highlight the growing interest in evaluating reasoning and compositional capabilities beyond isolated function synthesis. However, such benchmarks still do not fully capture repository-level software evolution and realistic pull-request-based development workflows.

SWE-bench~\cite{jimenez2023swe} is a curated collection of 2,294 issue–pull request pairs extracted from 12 popular Python repositories, designed to assess whether models can automatically resolve software issues by producing the corresponding reference pull request. Each instance in SWE-bench includes the textual description of the issue and the validated code patch, and correctness is assessed by executing the project test suites. This benchmark represents an important step toward realistic evaluation scenarios compared to earlier synthetic code-generation datasets; however, it remains limited in scale (only a few thousand examples from 12 repositories) and is largely monolingual (Python). 

PR4Code broadens both the scope and diversity of curated pull-request datasets. It collects more than 13,000 real pull requests from both Java and Python projects, each enriched with associated commits, NL descriptions, and metadata, with the goal of supporting large-scale, multi-language evaluation of code generation techniques. PR4Code covers heterogeneous types of development activities, including bug fixes, new features, and refactoring across two distinct language ecosystems, thus providing a broad and diverse benchmark in terms of task difficulty and content. Moreover, PR4Code is primarily intended for evaluation rather than large-scale training, preserving its role as an independent testbed to evaluate model generalization.

Beyond issue-centered benchmarks, datasets with large collections of commits and code diffs that can be used to train models and evaluate code understanding and automated bug-fixing capabilities are available. A representative example is MegaDiff~\cite{monperrus2021megadiff}, a corpus of more than 663,000 Java code differences extracted from repositories (largely derived from Defects4J~\cite{10.1145/2610384.2628055}) through strict filtering based on commit messages and diff size. MegaDiff supports tasks such as commit comprehension, fault localization, and automated program repair, and offers a large number of changes (all in Java) categorized by magnitude. However, datasets such as MegaDiff typically do not preserve the full structure of a pull request (e.g., multiple commits or rich textual descriptions), as they mainly focus on isolated diffs paired with commit messages. PR4Code, instead, retains the complete PR context: not only the final diff, but also the sequence of commits, the descriptive text at both PR and commit level, and additional metadata (e.g., authorship, timestamps, links to issues). This richer context makes PR4Code more suitable for simulating realistic code integration scenarios, where a model must understand and generate changes that are coherent with both a natural-language description and a development history.

Recent evaluation methodologies{~\cite{Wu2025RepoMasterEval}} {~\cite{jimenez2023swe}} have also emphasized the importance of large-scale, realistic benchmarks that go beyond isolated code snippets, incorporating execution-based validation, repository-level context, and task-oriented evaluation metrics. These approaches aim to better capture the complexity of real-world software development and provide more reliable assessments of model capabilities at scale.

Other datasets concentrate on very simple bug-fix patches, often at a function or single-statement level. For example, ManySStuBs4J~\cite{Karampatsis_2020} provides 153,652 single-statement bug fixes extracted from 1,000 Java projects, each labeled according to one of 16 frequent bug templates. This corpus is valuable for evaluating automated program-repair techniques on simple and repetitive defects, but it is highly homogeneous, being limited to Java, involving minimal code changes, and often accompanied only by short commit messages. In contrast, PR4Code includes pull requests of widely varying size, ranging from a few-line corrections to larger feature implementations affecting tens of lines and multiple files, and it exhibits substantial variability in the quality and length of textual descriptions, from short titles to detailed PR narratives. This ensures greater heterogeneity in both code and natural language, requiring evaluated models to cope with vague or informal descriptions and to generate or explain non-trivial code changes. Similarly to ManySStuBs4J for Java, the FixJS dataset~\cite{9796277} was proposed in the JavaScript domain, extracting approximately 300,000 before/after function pairs from about two million commits, while preserving original commits and modified files to provide rich context. Although FixJS represents a significant advance in scale and detail for training automated repair models, it remains focused on single-function changes within one language, whereas PR4Code targets multi-file, cross-artifact modifications that are more typical of real collaborative pull requests.

A complementary line of work investigates the interaction between AI agents and software development in the wild. 
Recent works \cite{hong2024metagpt,qian2024chatdev} on autonomous coding agents have further explored the ability of LLM-based systems to iteratively plan, generate, and refine code within realistic development environments, often combining multiple roles (e.g., planner, coder, reviewer) and interacting with external tools and repositories. These approaches highlight both the potential and the current limitations of agent-based software development, especially when dealing with complex, multi-step and multi-file tasks.

The recent AIDev dataset\cite{li2025riseaiteammatessoftware}, for example, captures more than 456,000 pull requests generated by autonomous agents (e.g., Codex- or Copilot-based systems) across approximately 61,000 GitHub repositories, together with rich metadata on authorship (human vs. AI), review timelines, code review outcomes, and merge decisions. AIDev provides a unique observational perspective on how AI-generated contributions are received and how they differ from human-authored PRs in terms of complexity and quality. Although AIDev is not structured as a controlled evaluation benchmark, it highlights the importance of studying realistic development artifacts and the limitations of purely synthetic benchmarks such as SWE-bench for answering questions about in-the-wild behavior. In this respect, PR4Code shares the emphasis on realism and on pull requests as the central unit of analysis, but differs in its objective: PR4Code offers a fixed, curated set of historical PRs to support repeatable experimental evaluation, whereas AIDev serves as a living corpus for empirical observation. Both, however, stress the relevance of including rich metadata and heterogeneous file types, as real PRs typically involve not only source code but also tests, comments, and documentation.

PR4Code positions itself within the landscape of recent datasets and benchmarks for code generation and modification as a complementary and necessary resource. Compared to commit-level corpora and small-patch datasets, it provides a higher-level, pull-request-centric view of software change, including multiple commits and heterogeneous artifacts. Compared to issue-focused benchmarks such as SWE-bench, it offers greater diversity in terms of projects, programming languages (Java and Python), types of development activities, and styles of natural-language descriptions. By capturing realistic, context-rich pull requests, it provides the research community with a robust and realistic testbed for assessing the capabilities of LLMs and AI agents in feature-level software development.

\section{Dataset Construction Methodology} \label{sec:methodology}
\begin{figure*}[h]
    \centering
    \includegraphics[width=0.8\linewidth]{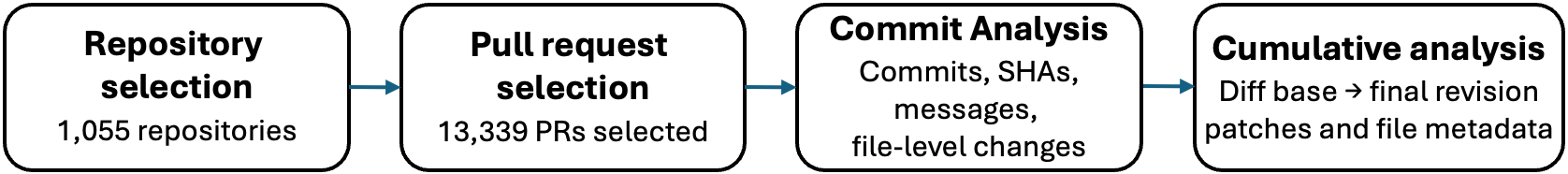}
    \caption{Methodology used to construct the \approach{} dataset}
    \label{fig:methodology}
\end{figure*}

The construction of the {\approach} dataset follows four sequential steps, as shown in Figure~\ref{fig:methodology}. The first step selects GitHub repositories that are relevant and remain active in terms of code changes and pull request creation. The second step selects closed pull requests that encapsulate feature development, bug fixing, or incremental improvements to the project. The third step analyzes the set of commits associated with each pull request. Finally, the fourth step analyzes the cumulative effect of the commits included in the same pull request. The output of this process is a curated and publicly available dataset of pull requests associated with the development of features extracted from relevant repositories.  

\subsection{Repository Selection}
We selected all GitHub repositories with significant popularity ($>$\textit{500 GitHub stars}), a threshold chosen to ensure a certain level of community adoption, relevance, and maturity, with recent pull requests created after the training cutoff date of several LLMs (after \textit{October 3, 2023}, which is the cutoff date of ChatGPT 4o), in order to reduce the risk of data leakage and ensure that the selected PRs are unlikely to be included in the training data of such models. We further restricted the selection to repositories having as primary language either \textit{Java} or \textit{Python}, the two most used languages in GitHub~\cite{GitHut2.0}.

This results in 1,055 repositories selected.

\subsection{Pull Request Selection}
For each selected repository, we retrieved and analyzed all the \textit{closed PR}s, that is, PRs that represent completed development activities. 
The goal of the analysis is to identify PRs representing feature development, bug fixing, or incremental improvement of the code, each accompanied by a description of the change, an essential 
component for constructing realistic input prompts for AI-based code generation.
To this end, a PR is selected in {\approach} if it satisfies the following conditions:

\begin{itemize}
    \item \textbf{Kind of development activity mentioned in the PR:} we select only the PRs whose title or body include at least one of the following keywords \texttt{requirement}, \texttt{feature}, \texttt{story}, \texttt{add}, \texttt{implement}, \texttt{enhance}, \texttt{remove}, \texttt{upgrade}, \texttt{change}, \texttt{create}, \texttt{resolve}, \texttt{develop}, \texttt{fix}, \texttt{update}, \texttt{support}, \texttt{merge}, \\ \texttt{refactor}, \texttt{improvement} or \texttt{build}.

    \item \textbf{Sufficient text:} the combined length of the PR title and body must be of at least \textit{100 characters}, ensuring a sufficiently detailed description of the developer’s intent; this threshold was introduced to exclude trivial or uninformative PRs (e.g., very short descriptions such as “fix bug”), which do not provide enough context for constructing meaningful input prompts, as observed during a preliminary inspection of PR descriptions of varying lengths.
    
    \item \textbf{Semantic branch:} to select PRs that actually represent feature development, bug fixing, or incremental improvement of the code, we rely on branch naming conventions~\cite{PullRequestNamingGuide, BestPracticesForNamingGitBranches}. That is, we retrieve the Git branch associated with the PR, and select only those PRs whose name of the associated branch includes a keyword among \texttt{feature}, \texttt{feat}, \texttt{story}, \texttt{task}, \texttt{fix}, \texttt{chore}, \linebreak \texttt{enhancement}, \texttt{improvement}, or \texttt{spike}.

\end{itemize}

\noindent We selected a total of 13,339 PRs, by applying the criteria above, 4,508 from Java projects and 8,831 from Python projects.

\subsection{Commit Analysis}
We retrieved the information about the set of commits associated with each PR, to precisely understand how the project was changed because of the PR. 
For each commit, we extracted the \textbf{unique SHA identifier} and the \textbf{timestamp} (author date) of the commit to enable the reconstruction of the development activities; the \textbf{commit message} describing the implemented change; %
the \textbf{list of modified files}; and, for each file, the number of added lines (\texttt{additions}), deleted lines (\texttt{deletions}), and total changes (\texttt{additions + deletions}). We also counted the  \textbf{textual references} within the commit message, aggregating URLs, markdown links, and issue identifiers. These references are useful to discriminate self-contained commit descriptions (i.e., descriptions with no references) and commit descriptions that require a broader context to be interpreted (i.e., the descriptions of the referred entities). 
{\approach} stores all commit-level information alongside the corresponding pull request, allowing cross-referencing between commit history, textual descriptions, and code modifications.

\subsection{Cumulative Analysis of Code Changes}
To reconstruct the full set of modifications done by developers to implement a PR, we computed the  diff between the project revision available before and after the completion of the PR. Specifically, we compared the \textbf{base snapshot} of the PR to the \textbf{final snapshot} associated with the merged changes. This comparison yields the complete set of changes introduced throughout the lifecycle of the pull request. 

For each modified file, we retrieved its \textbf{path}; its modification \textbf{status} (\textit{added}, \textit{modified}, or \textit{removed}); the number of lines \textbf{added}, \textbf{deleted}, and total \textbf{changed}; the textual \textbf{patch} (line-by-line diff); and
the URLs for direct file retrieval (\texttt{blob\_url}, \texttt{raw\_url}, \texttt{contents\_url}). This metadata provides a detailed and faithful representation  of both the size and complexity of the modifications applied within the scope  of each PR.

\smallskip

The dataset, the scripts used to build it, and the result of the selection process are available online~\cite{repository}.

\section{The \approach Dataset} \label{sec:approach}
The {\approach} dataset is organized as a hierarchical collection of repositories and PRs. For each PR, \approach includes the files modified by the PR, both in the version before and after the changes have been applied, together with a JSON file containing metadata about the changes. Metadata is also aggregated at the level of the entire dataset, distinguishing between Java and Python projects. The metadata includes all the information extracted during the dataset construction process, as described in Section~\ref{sec:methodology}.

\begin{figure}[h]
\centering
\includegraphics[width=\linewidth]{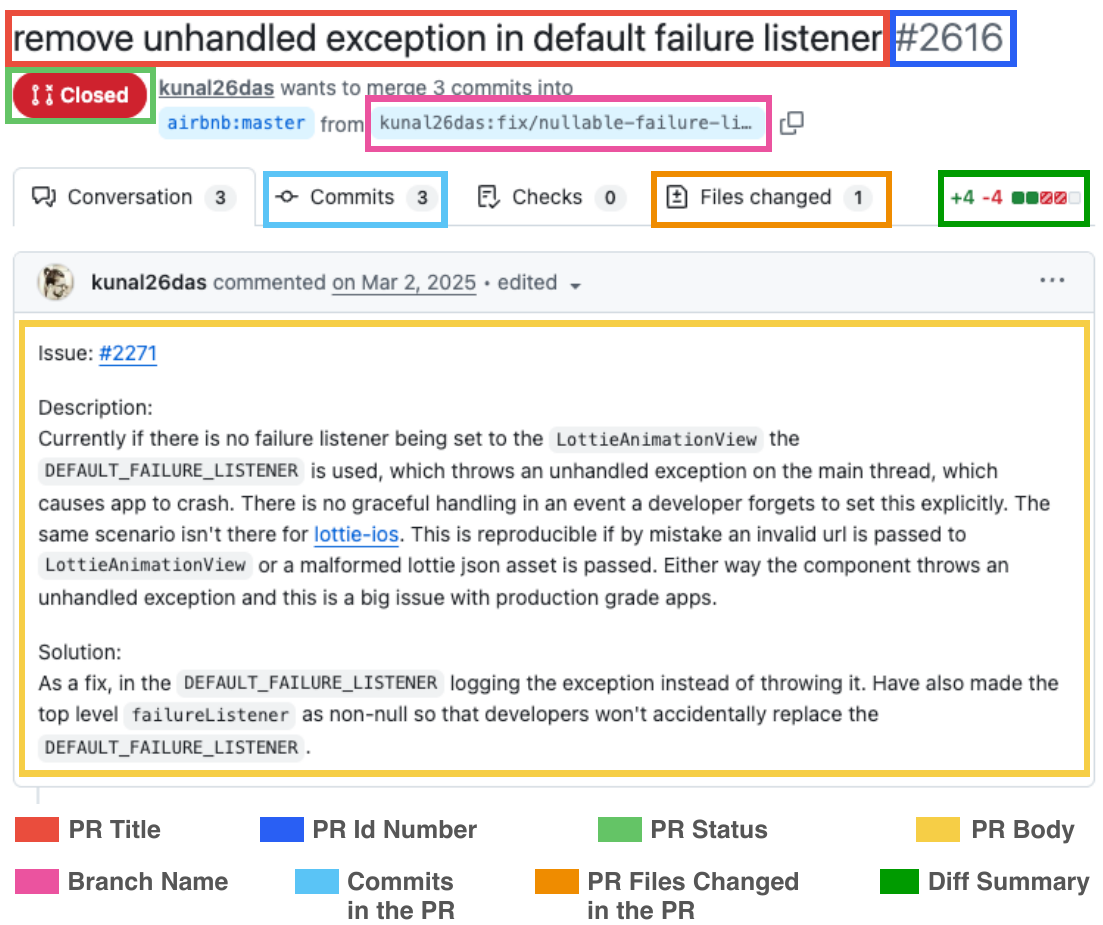}
\caption{Example of a pull request included in PR4Code.}
\label{fig:pr_example}
\end{figure}

To illustrate the structure and content of PR4Code, Figure~\ref{fig:pr_example} shows an example of a pull request as it appears on GitHub. The figure highlights the main elements considered during dataset construction, including the PR title and identifier, the PR status, the associated branch name, the natural-language description (body), the number of commits in the PR, the number of files changed in the PR, and the diff summary associated with the changes.

These elements are systematically extracted and stored in {\approach} as structured metadata, enabling the reconstruction of the development task from its textual description to the corresponding code modifications. The example also illustrates the realistic nature of the PRs included in the dataset, which often involve detailed descriptions with references or links to external artifacts, multiple commits and modifications affecting several files.

In the following, we describe the content and structure of the dataset in more detail.%

\subsection{Descriptions in PRs and Commits} \label{sec:prdec} %
We analyzed the descriptions provided by developers within PRs and commits, as this information is essential for defining realistic and context-aware prompts for AI-based code generation. In particular, we computed the length and content of 4,508 PRs and 14,639 commits from Java repositories, and 8,831 PRs and 31,695 commits from Python repositories. Below, we discuss the main findings of this analysis from multiple perspectives.

\subsubsection{Keyword frequency} \label{Keywordfrequency}
\begin{figure*}[!h]
    \centering
    \includegraphics[width=0.8\linewidth]{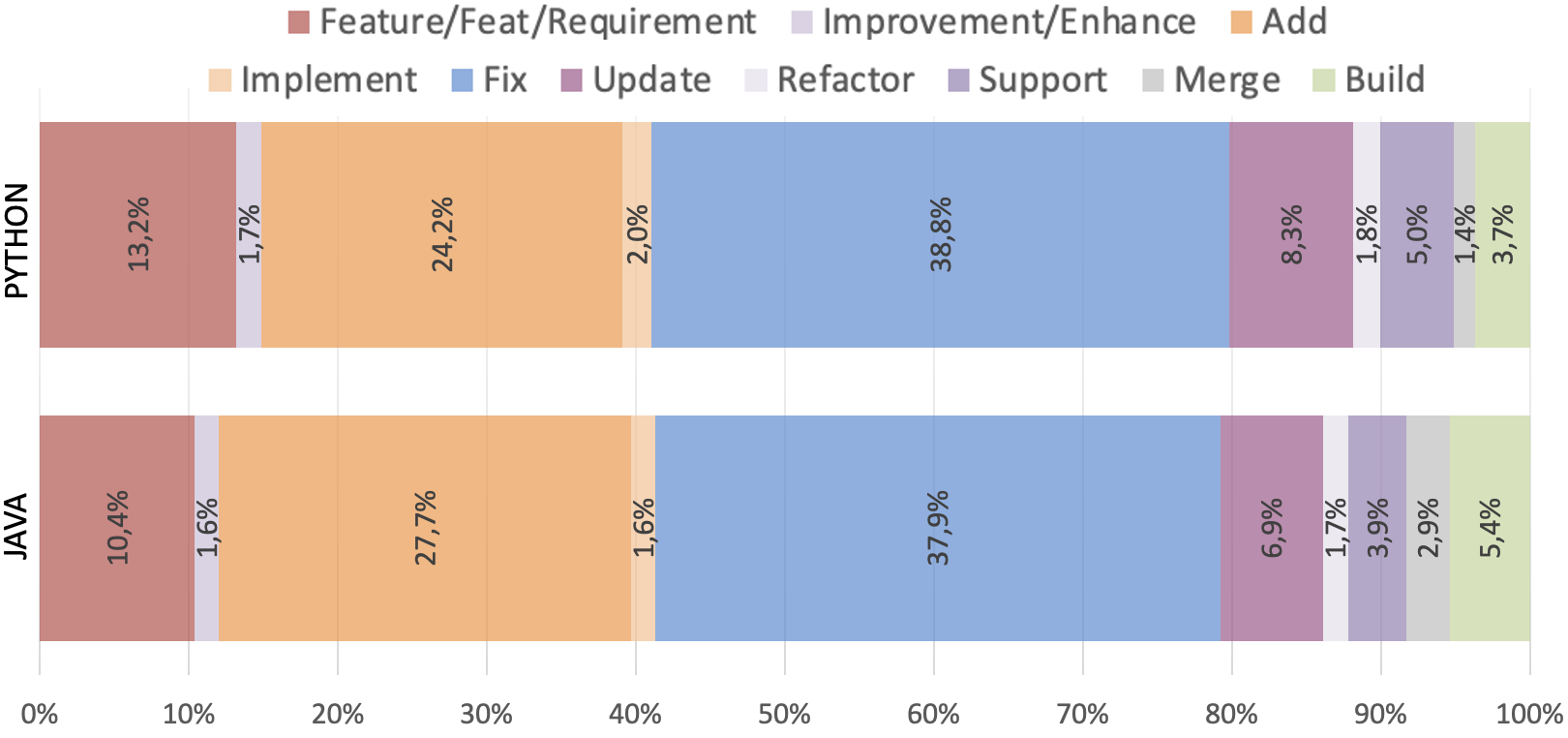}
    \caption{Frequency of keywords in PR titles and bodies for Java and Python projects, highlighting the most common types of development activities captured in \approach.}
    \label{fig:keywordinPRTitleBody}
\end{figure*}

To characterize the purpose of PRs, we analyzed the occurrence of the keywords used for PR selection in the title and body of PRs. Figure~\ref{fig:keywordinPRTitleBody} shows the results. The goal of this keyword-based analysis is not to measure the importance or complexity of PRs, but rather to characterize the diversity of development activities represented in the dataset. PR complexity is analyzed separately through other dimensions, such as the number of commits, modified files, and size of code changes.

To assess the reliability of the keyword-based characterization reported in Figure{~\ref{fig:keywordinPRTitleBody}}, we manually inspected a sample of 300 PRs balanced across Java and Python projects. Since a PR may contain multiple keywords, we validated individual keyword occurrences by checking whether they were semantically coherent with the actual intent of the PR. Overall, we validated 888 keyword occurrences, of which 78.49\% were classified as coherent. The highest coherence values were observed for more specific terms such as \textit{fix} (92\%) and \textit{refactor} (94\%), while more generic keywords such as \textit{update} (61\%) or \textit{merge} (39\%) were more prone to ambiguous usages. These results suggest that the keyword distribution provides a meaningful, although not perfect, high-level characterization of development activities.

Consistently with this validation, the keywords \textit{add} and \textit{fix} are the most frequent ones, indicating that adding new functionality and fixing existing behavior are among the most common development activities captured in the PRs. This suggests that {\approach} reflects common and representative software evolution tasks, rather than niche or highly specialized changes. Requirement implementation and software updates are also relatively common, while the remaining keywords occur less frequently.
The similarity between Java and Python further indicates a consistent way in which developers describe code evolution across different language ecosystems.

\subsubsection{Length and structure of textual descriptions} 
\begin{figure}[h]
  \centering
    \includegraphics[width=0.3\textwidth]{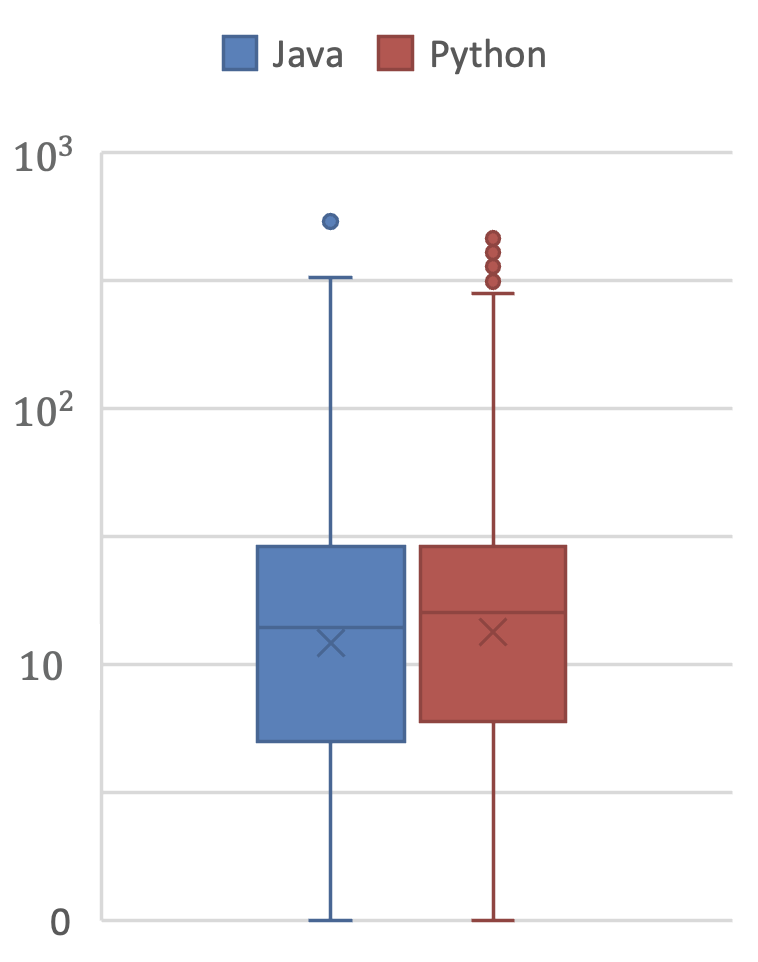}
    \caption{Distribution of the number of lines in PR descriptions.}
    \label{fig:PRlines}
\end{figure}

\begin{figure}[h]
    \centering
    \includegraphics[width=0.3\textwidth]{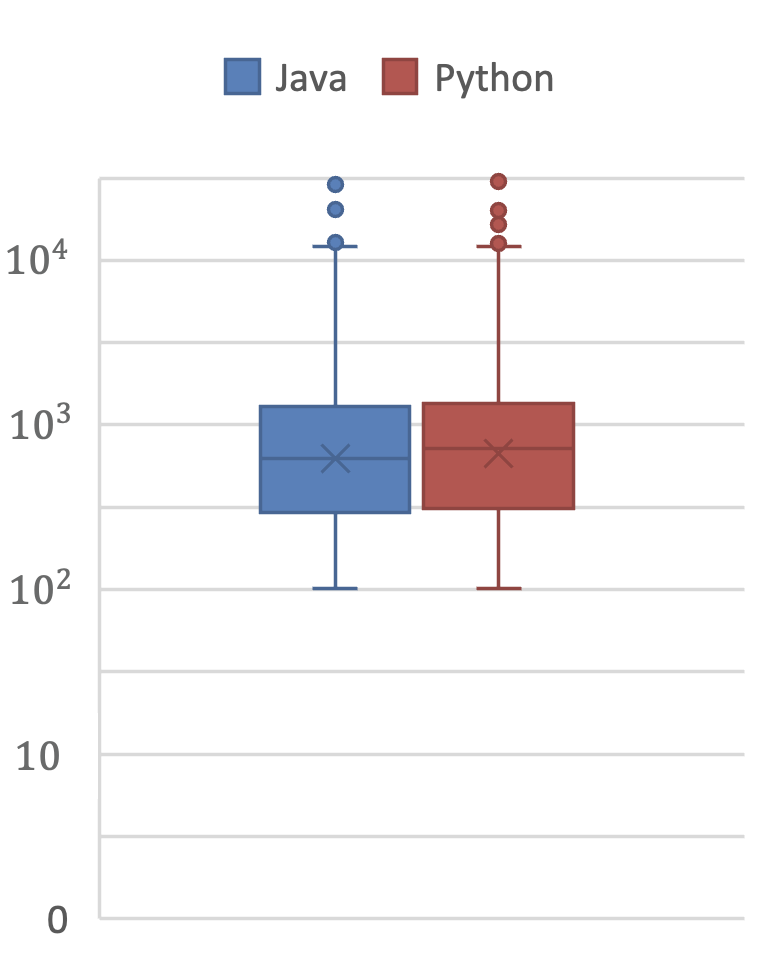}
    \caption{Distribution of the number of characters in PR descriptions.}
    \label{fig:PRchars}
\end{figure}

Figures~\ref{fig:PRlines} and~\ref{fig:PRchars} show the size of PR descriptions, considering the number of lines and characters. {\approach} includes a wide range of real cases, from extremely short to extremely long descriptions (e.g., thousands of lines and characters) revealing substantial variability in how developers document their work. This variability reflects the diversity of real-world development practices, where descriptions may range from minimal summaries to detailed narratives including context and rationale. %

For Java projects, the PR body has an average length of 19.3 lines (median 13), while for Python projects, the average length is 21.3 lines (median 15).
In terms of character count, Java PR bodies contain on average 986 characters (median 577), compared to 1,040 characters (median 669) for Python. The most verbose descriptions exceed 30,000 characters. %

Titles are substantially more compact, with an average of 51 characters in Java and 48 in Python, reflecting a consistent tendency toward concise summarization practices across languages. 

PR descriptions and titles report information about the nature and goal of changes, providing valuable context for assessing AI agents tasked with planning and generating code for full feature implementations.

\begin{figure}[h]
    \centering
    \includegraphics[width=0.3\textwidth]{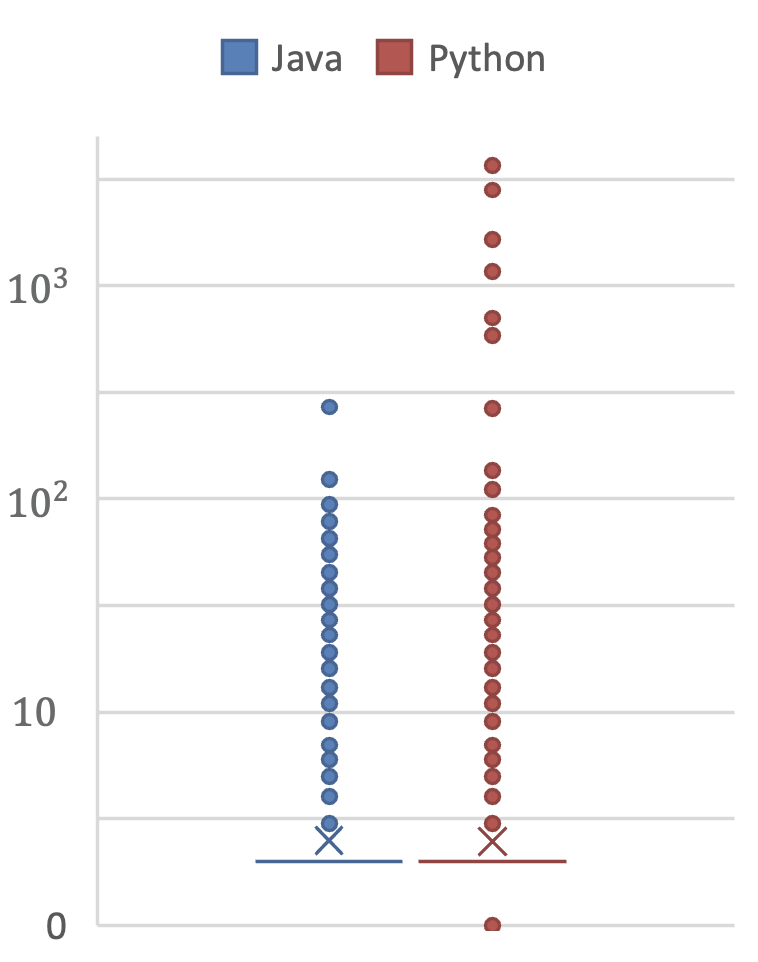}
    \caption{Number of lines per commit description.}
    \label{fig:commitlines}
\end{figure}

\begin{figure}[h]
    \centering
    \includegraphics[width=0.3\textwidth]{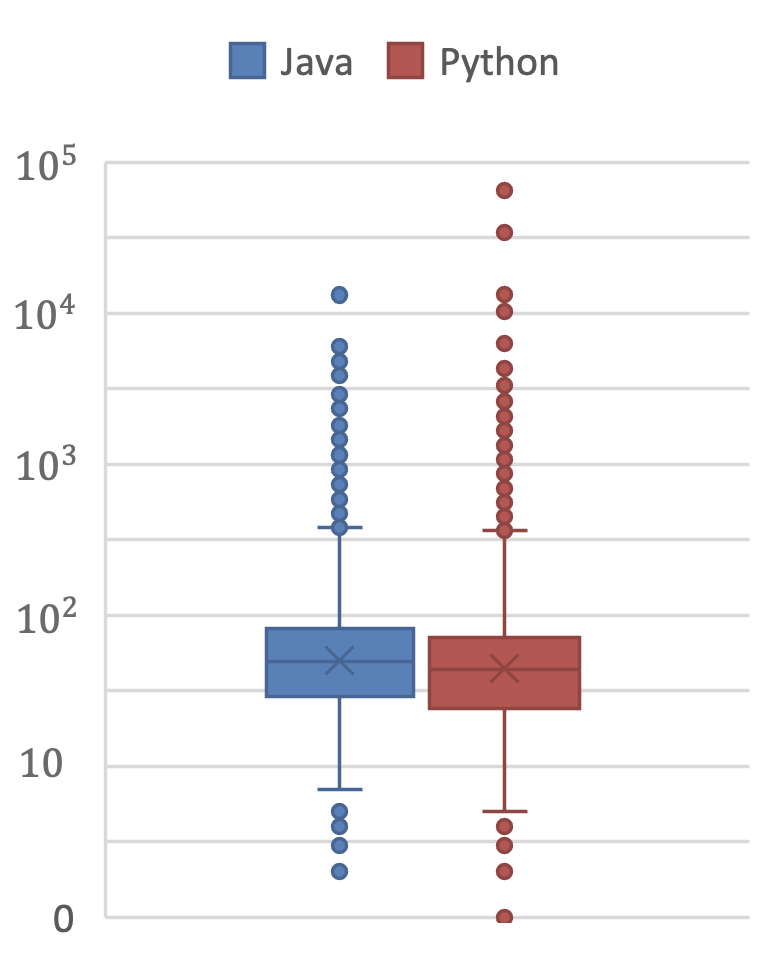}
    \caption{Number of characters per commit description.}
    \label{fig:commitChars}
\end{figure}

Figures~\ref{fig:commitlines} and~\ref{fig:commitChars} show the size of commit descriptions, considering the number of lines and characters. Not surprisingly, commit descriptions are shorter than PR descriptions, as they are intended to document fine-grained and incremental changes. This distinction reinforces the hierarchical nature of development information in {\approach}, where high-level intent is captured at the PR level and low-level actions at the commit level.
The average message length is 86 characters in Java (median 48) and 105 in Python (median 43), typically spanning 1 or 2 lines. %
Yet, these descriptions document individual code-level changes that can be used to train and assess AI agents implementing features. 

Overall, these quantitative figures indicate that PR descriptions typically provide an informative summary of the implemented changes, although with considerable variability in length and detail. In contrast, commit messages tend to offer a more concise and operational view of individual changes, highlighting the complementary roles of PR- and commit-level descriptions.

\subsubsection{Hyperlinks and cross-references} Sometimes descriptions are not self-contained, since they may include references to external sources. 
Expressions that automatically close issues (\textit{Fixes \#...}, \textit{Closes \#...}, \textit{Resolves \#...}) occur in 1,201 Java and 2,402 Python PRs, confirming the diffusion of GitHub automation syntax. Hyperlinks are predominantly located in the body of pull requests, where developers provide contextual information about the requirements and the functionalities to be implemented. In particular, 2,487 out of 4,508 Java PRs and 4,576 out of 8,831 Python PRs contain at least one external link.

The diffusion of hyperlinks and cross-references suggests that PRs typically consist of structured text, not just plain text. Overall, 55\% of Java PRs and 52\% of Python PRs include at least one reference to an external artifact. 
This indicates that fully understanding the nature of a PR may require (transitively) accessing several other sources, challenging, for instance, the construction of prompts for AI assistants.

Our analysis of template usage reveals a consistent adoption of standardized textual structures across PRs. 
In particular, markdown checklists (\texttt{- [ ]}) appear in 2,431 out of 4,508 Java PRs and 5,106 out of 8,831 Python PRs, indicating a broader adoption of structured templates in Python projects. Collaboration markers such as \textit{Signed-off-by} and \textit{Co-authored-by} are also more frequent in Python, reflecting more explicit authorship tracking practices. In particular, Signed-off-by tags occur in 529 Java PRs and 492 Python PRs, while Co-authored-by markers appear in 147 Java PRs and 495 Python PRs.
The use of structured information may simplify automated analysis and text understanding, offering additional analysis opportunities.

\subsection{Characteristics of the PRs}
We analyzed the structure and content of PRs, considering their commits, file and line modifications, file types, and the occurrence of tests and comments in the changes.

\subsubsection{Commits per PR} 
\begin{figure}[h]
  \centering
    \includegraphics[width=0.3\textwidth]{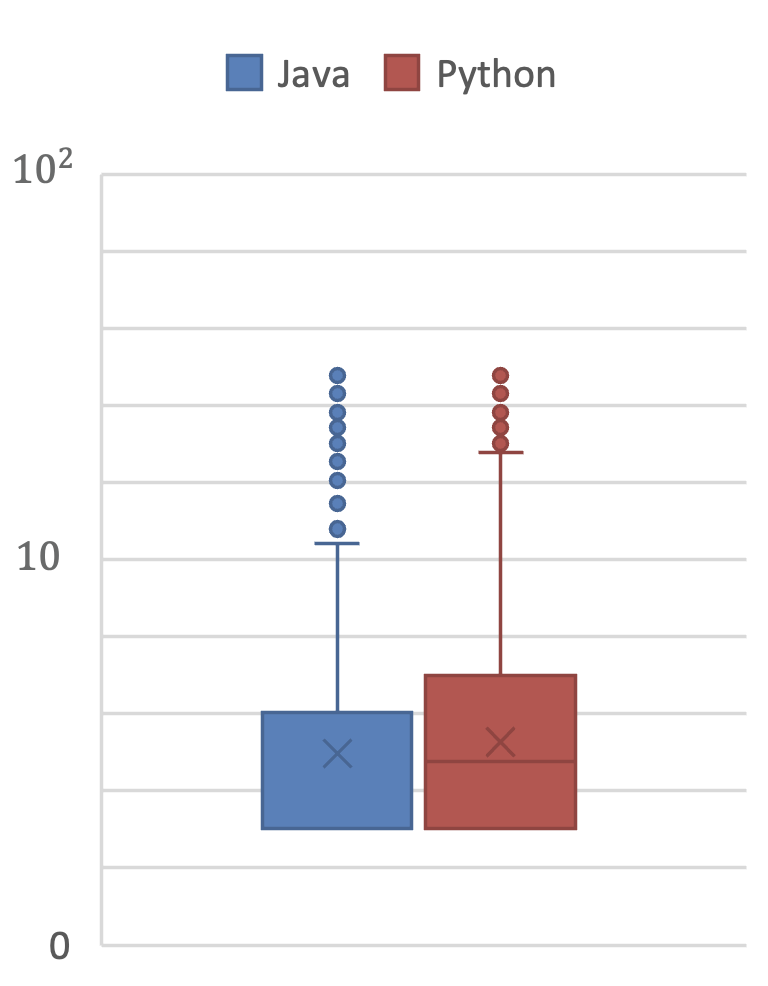}
    \caption{Distribution of the number of commits per PR.}
    \label{fig:commitPerPR}
\end{figure}

Figure~\ref{fig:commitPerPR} shows the number of commits per PR. In both Java and Python the number of commits per PR is usually small, indicating that most development tasks are implemented through a limited sequence of coherent changes. This suggests that many feature implementations are relatively structured and manageable in terms of development steps. In Java, the average number of commits per PR is 3.25 (median 1), with 54\% of the PRs containing a single commit and 75\% of the PRs no more than three. In Python, the average number of commits per PR is 3.59 (median 2), with 47\% of the PRs containing a single commit and 75\% of the PRs no more than four. The noticeable gap between mean and median values indicates a skewed distribution, where most PRs contain a few commits, while a limited number of PRs involve substantially larger commit sequences. Manual inspection of several high-commit PRs revealed that these cases are often associated with large-scale development activities, such as framework migrations, rendering-engine refactorings, infrastructure updates, and feature branches that underwent multiple review-driven revisions before merge.
Only 7.7\% and 10.5\% of the PRs in Java and Python projects exceed eight commits (with a maximum of 30), indicating that most PR implementations are concise and composed of few logically distinct changes. As a consequence, the median better reflects the typical PR size in the dataset, whereas the mean is influenced by these less frequent but more complex development scenarios. This insight is helpful for the design of AI-based code generators that must calibrate the magnitude of suggested changes according to the description of the code-generation task.

\begin{figure}[h]
    \centering
    \includegraphics[width=0.3\textwidth]{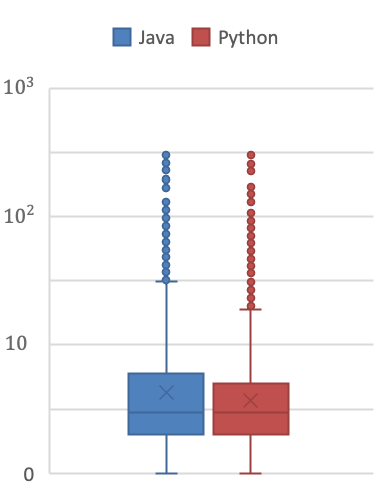}
    \caption{Distribution of the number of files modified per PR.}
    \label{fig:filePerPR}
\end{figure}

\subsubsection{Files modified per PR} Figure~\ref{fig:filePerPR} shows the number of files modified by PR.
In Java, PRs modify on average 3.47 files and in Python 4.26 files. Most PRs (75\% for Java projects and 70\% for Python projects) affect only one or two files, while a few involve hundreds, typically due to refactoring or large merges. The fact that many of the changes are localized to a few files and distributed across a few commits only suggests that many features can be added to projects by implementing fairly localized changes. This also indicates that a substantial portion of feature  implementations is modular and contained, which is particularly relevant for  AI-assisted code generation techniques that must reason about localized modifications.

\subsubsection{Lines added, removed, and changed}
\begin{figure} [h]
    \centering
    \includegraphics[width=0.45\textwidth]{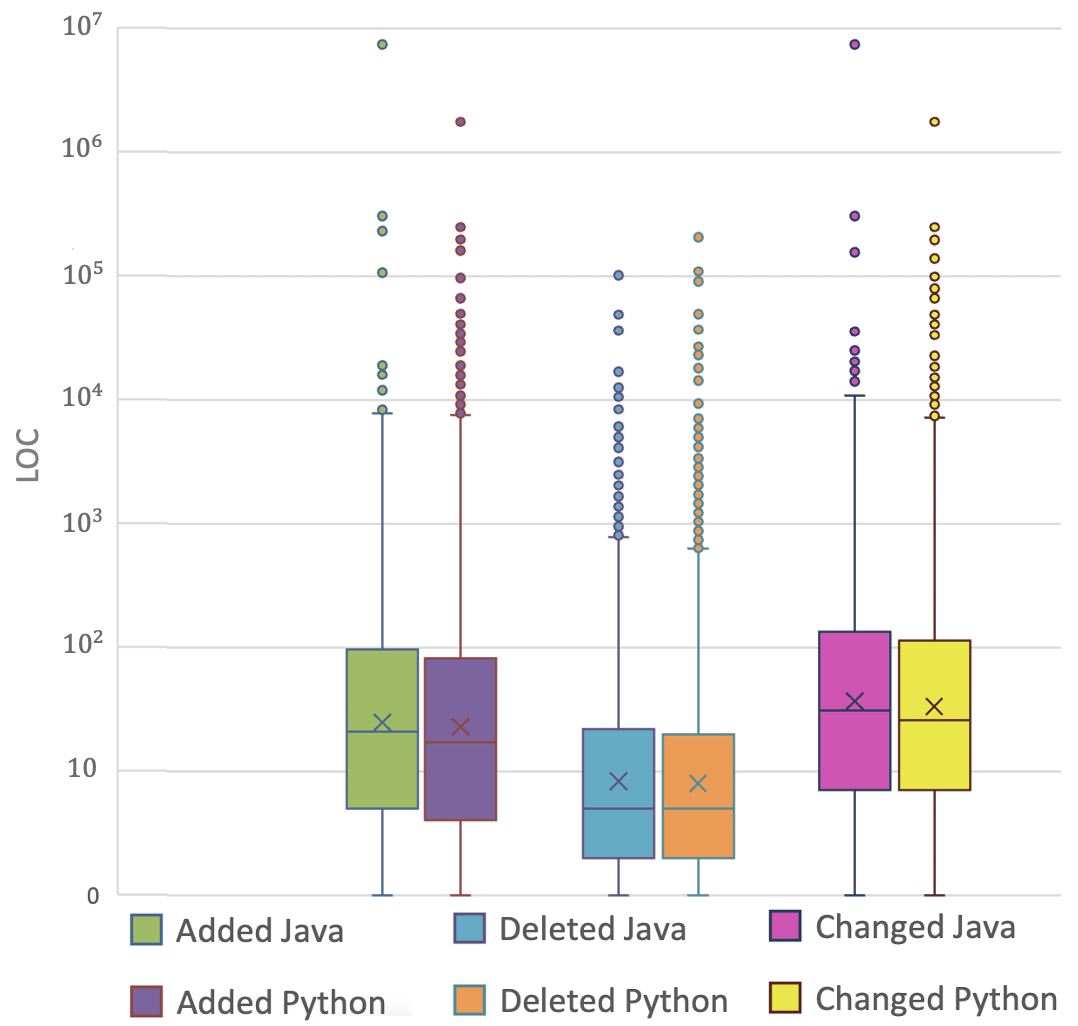}
    \caption{Distribution of added, deleted, and modified lines per PR.}
    \label{fig:FilesModifiedPerPR}
\end{figure}

Figure~\ref{fig:FilesModifiedPerPR} shows the number of lines added, removed, and changed per PR. The distribution is extremely skewed, with several outliers, reflecting the coexistence of small-scale maintenance commits and large-scale restructuring operations. This variability is particularly relevant for evaluation purposes, as it allows assessing model performance across tasks of different sizes and complexity levels. 

Although changes are localized in a few commits and files only, the magnitude of changes within files is quite spread, with several PRs modifying more than 100 lines of code, which is a challenging magnitude for AI-based techniques.

\subsubsection{Type of modified files} Not all changes are limited to source code files. In particular, only 44.4\% of Java PRs and 50.6\% of Python PRs exclusively modify language-specific source files (i.e., \texttt{.java} files for Java projects and \texttt{.py} files for Python projects). We found that 23.8\% of PRs in Java projects and 25.6\% of PRs in Python projects modify both language-specific source files and other types of files. Surprisingly, a relevant portion of PRs, 31.8\% of PRs in Java projects and 23.5\% of PRs in Python projects, do not modify any Java or Python file. This highlights the prevalence of changes targeting non-code artifacts, which are essential components of modern software systems. %

\begin{figure} [h]
    \centering
    \includegraphics[width=1\linewidth]{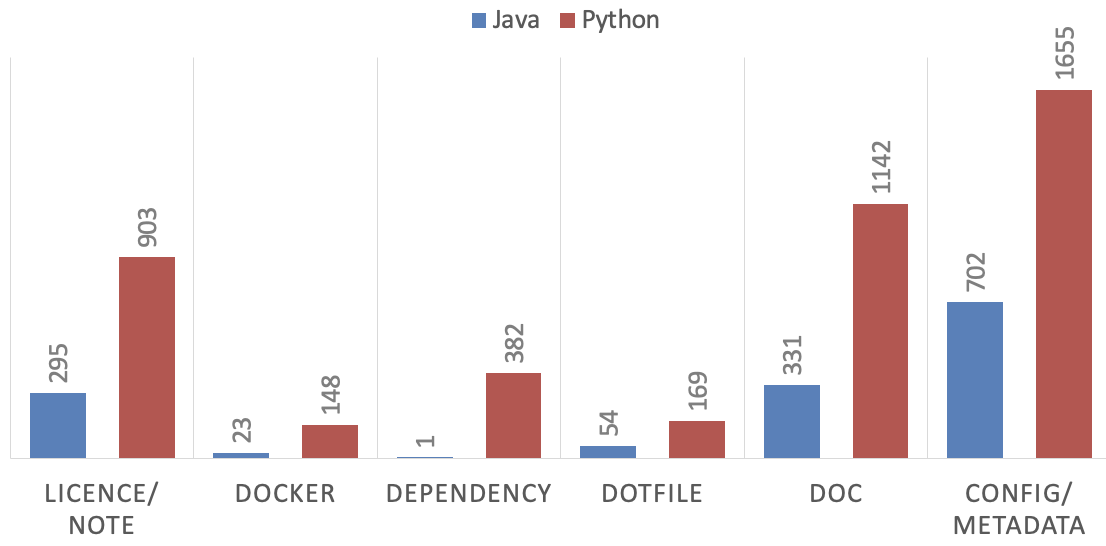}
    \caption{Number of PRs that modify non-code artifacts.}
    \label{fig:PRNonSourceCodeFiles}
\end{figure}

We analyzed the nature of changes not targeting source-code files. Figure~\ref{fig:PRNonSourceCodeFiles} shows the results. Changes mainly target configuration and metadata files (e.g., \texttt{README.md}, \texttt{LICENSE}, \texttt{CHANGELOG.md}), followed by documentation, licensing, and dependency-related artifacts,  highlighting the importance of non-code elements in real-world development. Changes to these files are more frequent in Python projects than in Java projects. %

This result is particularly relevant for AI‑based code generation, showing that effective feature development requires the ability to handle multiple types of files rather than merely generating source code. The {\approach} dataset provides a wide variety of representative cases that can be used to assess future AI-based solutions.

\subsubsection{Presence of test files} Indeed, changes may affect not only the components of the applications hosted in the repositories, but also their test files. To estimate how often changes to tests occur, similarly to other studies~\cite{miranda2025test, levin2017co}, we measured the percentage of changes affecting files whose names or hosting folders include the string \texttt{test}. We discovered that 38.2\% of PRs in Java and 33.4\% of PRs in Python include changes to at least one test file. These findings confirm that PR implementations also require the evolution of test files, as also reported in previous studies~\cite{miranda2025test, levin2017co}.   

\subsubsection{Comment-related changes} Code evolution requires not only evolving the code, but also evolving the comments that describe the code. 
We detected code comment modifications as changes to comment lines defined with  \texttt{//}, \texttt{/*}, or \texttt{\#}. We discovered that nearly half of the changes affect comments (49.2\% of PRs in Java, 54.6\% of PRs in Python), while comment-only changes are rare (below 1\%). These results highlight the importance of co-evolving code and documentation, and indicate that AI-based solutions must be capable of generating and updating comments alongside source code. 

\subsection{Characteristics of Commits} 

We briefly summarize here data about commits. 

\subsubsection{Lines added, removed, and changed} Commits intuitively represent small changes that evolve the project in a semantically coherent way. Those units of change serve as the basic granularity that AI-based tools have to mimic when attempting to emulate the work done by developers. The average number of modified lines per commit is non-trivial; in fact, it amounts to 37.8 lines modified across 9.38 files for Java projects and 42.7 lines modified across 5.97 files for Python projects. 

\subsubsection{Type of modified files}
Similarly to PRs, the percentage of commits modifying source code files only amounts to 56.1\% and 50.8\% for Python and Java projects, respectively.
While mixed commits involving auxiliary files (e.g., configuration or documentation) account for about one-fifth. 
Between 26\% and 30\% of the commits modify non-code files only, such as build scripts or README documents. 
These results confirm that, although source code changes dominate, a significant portion of activity targets peripheral project elements. 
The higher share of code-only commits in Python reflects its simpler structure, whereas Java’s ecosystem often results in more non-code changes due to its complex build and deployment configurations.

\subsubsection{Presence of test files}
Also, for changes to test files, the commit-level statistics resemble the PR-level statistics, with 29.1\% of Java commits and 34.5\% of Python commits involving at least one test file. \\

To further characterize the differences between Java and Python pull requests, we performed a statistical comparison across the main quantitative dimensions of the dataset using the Mann–Whitney U test, a non-parametric statistical test suitable for skewed and non-normally distributed software engineering data. To complement statistical significance testing and avoid overinterpreting minimal differences in large datasets, we additionally computed Cliff’s Delta and Vargha–Delaney A12 effect sizes to quantify the magnitude of the observed differences. Table~\ref{tab:statistical_comparison} summarizes the results.

\begin{table*}[h]
\centering
\caption{Statistical comparison between Java and Python PRs across the main quantitative dimensions of PR4Code.}
\label{tab:statistical_comparison}
\begin{tabular}{lrrrrlr}
\toprule
Metric & Java median& Python median & p-value & Cliffs delta & Effect size & Vargha–Delaney \\
\midrule
Commits per PR & 1 & 2  &$<10^{-3}$  & -0.0662 & negligible & 0.4669\\
LOC added per PR & 20  & 16  & 0.01 & 0.0274 & negligible & 0.5137\\
LOC deleted per PR & 4  & 4  & 0.14 & 0.0154 & negligible & 0.5077\\
LOC changed per PR & 30  & 25  & $<10^{-3}$ & 0.0349 & negligible & 0.5175\\
PR title length & 49  & 46  & $<10^{-3}$  & 0.0780 & negligible & 0.5390\\
PR body length & 576  & 668  & $<10^{-3}$  & -0.0449 & negligible & 0.4775\\
PR body lines & 13  & 15  & $<10^{-3}$  & -0.0450 & negligible & 0.4775 \\
Commit message length & 48  & 43  & $<10^{-3}$  & 0.0854 & negligible & 0.5427 \\
Commit message lines & 1  & 1  & $<10^{-3}$ & 0.0209 & negligible &  0.5104\\
Files changed per commit & 2 & 1 & $<10^{-3}$ & 0.1031 & negligible & 0.5515\\
Source-only PRs & 0 & 1 & $<10^{-3}$ & -0.0622 & negligible & 0.4689\\ 
\bottomrule
\end{tabular}
\end{table*}

Although several comparisons resulted in statistically significant differences, all observed effect sizes are negligible according to both Cliff’s Delta and Vargha–Delaney A12 statistics. This indicates that, despite minor distributional variations, Java and Python PRs exhibit overall comparable characteristics in terms of commit structure, code churn, textual description size, and file modification patterns. The results therefore suggest that the dataset maintains a relatively balanced structure across the two language ecosystems, reducing the risk that subsequent analyses are disproportionately influenced by language-specific artifacts rather than by general software development behaviors.

The largest differences are observed for metrics such as files changed per commit and commit message length; however, even in these cases the effect sizes remain negligible. This further supports the view that the two subsets share broadly similar development dynamics despite small variations in coding and contribution practices.

\subsection{External Validation of Dataset Characteristics}

To further validate the construction of {\approach}, we compared its characteristics with both an existing repository-level benchmark and a random sample of GitHub pull requests. This comparison aims to assess whether the dataset reflects realistic pull-request-based development activities and whether the keyword-based selection introduces an artificial characterization of PR intent.

First, we compared {\approach} with SWE-bench~\cite{jimenez2023swe}, one of the closest existing resources for repository-level evaluation. As summarized in Table~\ref{tab:dataset_comparison}, SWE-bench contains 2,294 issue--PR pairs from 12 Python repositories and focuses on executable issue-resolution tasks, {\approach} includes 13,339 pull requests from 1,055 Java and Python repositories. Moreover, {\approach} covers a broader range of development activities, including bug fixes, feature additions, updates, refactorings, and changes involving heterogeneous artifacts such as source code, configuration files, documentation, and tests. This comparison highlights the complementary nature of {\approach}: rather than focusing only on issue resolution, it provides a broader and more diverse collection of pull-request-level development tasks.

\begin{table}[h]
\centering
\caption{Comparison between PR4Code and SWE-bench.}
\label{tab:dataset_comparison}
\begin{tabular}{lll}
\toprule
Dimension & SWE-bench & PR4Code \\
\midrule
\vspace{0.05cm}
Task unit & Issue PR pair & Pull request \\
\vspace{0.05cm}
Languages & Python & Java, Python \\
\vspace{0.05cm}
Repositories & 12 & 1,055 \\
\vspace{0.05cm}
Instances & 2,294 & 13,339 \\
\vspace{0.05cm}
Main focus & Issue resolution &
\begin{tabular}[t]{@{}l@{}}
Feature-level code\\
changes
\end{tabular}
\\
\vspace{0.05cm}
Repository diversity & Limited & Broad \\
\vspace{0.05cm}
Includes commits & Patch-level & 
\begin{tabular}[t]{@{}l@{}}
Commit sequences and\\
cumulative diff
\end{tabular}
\\
\vspace{0.05cm}
Non-code artifacts & Limited & Explicitly characterized \\
\bottomrule
\end{tabular}
\end{table}

Second, we collected a random sample of 300 closed GitHub pull requests, balanced across Java and Python projects and subject to the same temporal constraint adopted for {\approach}, but without applying the keyword, branch-name, or minimum-description-length filters. As summarized in Table~\ref{tab:random_pr_sample}, 19\% of the random PRs have an empty body and 27\% have a combined title and body shorter than 100 characters. This supports the need for a minimum textual-content threshold when constructing input prompts for AI-based code generation. At the same time, 71\% of the random PRs contain at least one of the selection keywords, indicating that terms such as ``add'', ``fix'', and ``update'' are common in general GitHub development activity rather than being artifacts introduced by our filtering process.

\begin{table}[h]
\centering
\caption{Characteristics of the random GitHub PR sample used for external validation.}
\label{tab:random_pr_sample}
\begin{tabular}{lrrr|r}
\toprule
Metric & Java & Python & All & PR4Code \\
\midrule
Median title+body chars & 436.5 & 823.5 & 661.0 & 683.5\\
Empty body & 27\% & 11\% & 19\% & None \\
Title+body $<$ 100 chars & 37\% & 18\% & 27\% & None \\
With selection keyword & 67\% & 74\% & 71\% & 100\%\\
With links/references & 50\% & 59\% & 55\% & 53\% \\
Median commits per PR & 1 & 1 & 1 & 2 \\
Median changed files per PR & 2 & 3 & 3 & 2\\
Median changed lines per PR & 69 & 119.5 & 95.5 & 28\\
\bottomrule
\end{tabular}
\end{table}

This comparison suggests that {\approach} selects pull requests with more explicit development intent and sufficient textual context, while preserving terminology and structural characteristics commonly observed in real GitHub pull requests. Together with the manual validation of keyword occurrences reported in Section~\ref{Keywordfrequency}, this provides additional evidence that the dataset construction is not solely heuristic-driven and that the selected PRs are suitable for realistic AI-based code generation studies.

\section{Discussion} \label{sec:discussion}
The \approach dataset includes PRs, commits, and code changes that offer a wide range of meaningful challenges for AI-based code-generation techniques. For instance, it includes changes of various sizes (from a few to hundreds of thousands of lines), targeting a variable number of elements (from a few to hundreds of code locations and files), that modify artifacts of heterogeneous nature (e.g., code, configuration files, comments, dependencies, licenses, etc.), and documented through natural-language descriptions at both PR 
and commit level. \approach can thus be used to address several research questions about software evolution and feature development. In light of the growing need to evaluate AI-based approaches on feature-level code changes, we discuss below representative usage scenarios.

\begin{itemize}
\item \textbf{Feature-level code generation}. {\approach} can be used to evaluate the ability of AI systems to implement complete software features starting from natural-language descriptions. In this scenario, the model receives the PR description and must generate the corresponding multi-file changes. This represents a realistic setting, where the model must interpret the developer's intent and produce coherent modifications across the codebase. The PRs in {\approach} cover a wide range of real-world feature implementations, varying in size, complexity, and description detail.

\item \textbf{Incremental development and commit modeling.} The availability of commit sequences enables the evaluation of incremental code generation. Models can be tasked with reproducing the evolution of a PR step by step, generating intermediate commits. This allows assessing whether the model can mimic realistic development workflows and maintain consistency across successive changes.

\item \textbf{Multi-file and heterogeneous artifact handling}. Since {\approach} includes changes affecting different types of artifacts (e.g., source code, configuration files, documentation, and tests), it can be used to assess the ability of models to handle heterogeneous development tasks. Implementing a feature often requires dealing with multiple file types, and nearly half of the PRs in {\approach} include non-code elements, enabling evaluation beyond traditional source code generation.

\item \textbf{Handling of cross-referenced and non self-contained descriptions.} PR descriptions are often not standalone, as they may include references to external artifacts (e.g., issues, links, or other PRs), as reported in Section~\ref{sec:prdec}. This introduces additional complexity, requiring models to interpret distributed and potentially incomplete information. {\approach} enables the evaluation of both prompt construction techniques and code generation approaches under these realistic conditions.
\end{itemize}

\noindent There are also limitations related to the {\approach} dataset. For instance, 
PR4Code focuses on pull requests from popular open-source Java and Python repositories selected using a popularity threshold. While this choice ensures realistic and non-trivial development scenarios, it may limit generalizability to less popular projects, other programming languages, or proprietary settings. 

Moreover, this selection process may introduce biases toward well-structured and actively maintained repositories, where development practices (e.g., clearer documentation, standardized workflows, and more informative PR descriptions) are more mature than in smaller or less active projects. As a consequence, the dataset may underrepresent noisy, poorly documented, or irregular development processes that are also common in real-world settings.

Similarly, the filtering criteria applied to pull requests (e.g., minimum text length and keyword-based selection) may bias the dataset toward PRs with more explicit and descriptive natural-language content, potentially excluding shorter or less structured PRs that still represent valid development activities. This could affect the evaluation of AI-based approaches, as models may be exposed to clearer input descriptions than those encountered in practice.

Moreover, keyword-based categorizations may not fully reflect the actual complexity or practical relevance of PRs, as rare but critical development activities may appear less frequently in the observed distributions. For this reason, the keyword analysis should be interpreted as a descriptive characterization of development activity types rather than as a direct measure of task importance or difficulty.

Nevertheless, Java and Python are widely used, and the dataset exhibits substantial diversity in project characteristics, PR sizes, and artifact types.

Restricting PR4Code to pull requests created after October 3, 2023 mitigates the risk of training data leakage for models with a known cut-off date, but does not eliminate it. Different LLMs may rely on partially undisclosed or continuously updated training corpora, and newer models may already include some PRs in their training data. The temporal constraint should therefore be interpreted as a best-effort mitigation; future studies should explicitly report the training horizon of evaluated models and, where feasible, apply additional leakage controls.

\section{Using \approach} \label{sec:usage}
{\approach} is designed not only as a curated dataset, but also as a practical resource for evaluating AI-based code generation techniques in realistic software engineering scenarios. In this section, we discuss how the dataset can be used in practice, outlining a typical usage workflow, possible application scenarios, and a small-scale experiment that demonstrates its applicability.

\subsection{Usage workflow}
{\approach} provides all the artifacts required to reconstruct real development tasks. For each PR, the dataset includes the natural-language description (title and body), the base version of the repository, the sequence of commits, and the cumulative code changes.
A typical workflow proceeds as follows. A PR is selected and its description is used as input to the model. The corresponding base version of the repository is retrieved using the provided commit identifier, enabling reconstruction of the initial state. The model is then tasked with generating the required changes. The output is compared against the reference implementation, either at the level of cumulative diffs or individual commits. The availability of structured metadata and repository links supports flexible and reproducible experimentation.

\subsection{Small-scale applicability experiment}
To further demonstrate the practical applicability of {\approach} and complement the descriptive analysis with evaluation-oriented results, we conducted a small-scale repository-level code generation experiment using an agentic pipeline based on the DeepSeek-R1:8B LLM. The objective of this experiment is to show that {\approach} can be operationally used to instantiate realistic pull-request-driven code generation scenarios involving repository-level modification, compilation, testing, and manual inspection of  generated patches. This experiment is intended to demonstrate dataset usability rather than to provide a statistically representative evaluation of model performance.

We selected 10 pull requests from {\approach} according to the following criteria: (i) the pull request described a localized modification that could be manually inspected, (ii) the repository could be compiled in isolation using Maven or Gradle, (iii) the pull request included either modified tests or executable tests related to the target functionality, and (iv) the change affected at most three files to enable detailed manual verification. The selected pull requests included bug fixes and small feature modifications involving real multi-file repository contexts.

For each pull request, we checked out the version of the repository immediately preceding the first commit associated with the PR. The agentic pipeline was then prompted to implement the feature from the pull request title and body. The pipeline was composed of four stages. First, a planning agent analyzed the pull request title and body to identify the intended modification and the repository files potentially involved in the change. Second, a context retrieval agent collected repository-level information related to the target files, including the original source code and surrounding project context. Third, a code generation agent produced modified source files starting from the original repository version and the retrieved contextual information. Finally, an evaluation agent automatically compiled the repository and executed the tests associated with the target functionality.

Each generated patch was evaluated according to four criteria:
\begin{enumerate}
    \item Successful compilation of the modified repository,
    \item Successful execution of the project tests,
    \item Manual verification that the requested change described in the pull request was effectively implemented,
    \item Manual identification of unrelated or unnecessary modifications outside the intended change scope.
\end{enumerate} The manual verification was performed by comparing the pull request description, the original developer patch, and the patch generated by the agentic pipeline. These criteria provide preliminary effectiveness metrics for repository-level code generation in realistic pull-request-driven development scenarios.

The experiment showed mixed results across the selected pull requests. Table~\ref{tab:repository_experiment} summarizes the selected PRs and the corresponding evaluation outcomes. Specifically, 6 out of 10 generated patches successfully compiled, while 4 out of 10 also passed the target tests associated with the pull request. Manual inspection showed that the requested modification described in the PR was correctly implemented in four cases, although the generated solution sometimes differed from the original developer implementation. This behavior is plausible since multiple semantically valid implementations may exist for the same pull-request requirement, especially in repository-level development scenarios involving refactoring decisions, alternative API usages, or different localization strategies for the modification. We also noticed three implementations that, although not passing the tests, were nearly identical to the developers' implementations. Furthermore, a very limited number of generated patches introduced unrelated or unnecessary modifications outside the intended change scope, suggesting that the pipeline generally remained focused on the requested repository changes.

\begin{table}[h]
\centering
\caption{Summary of the small-scale applicability experiment conducted with PR4Code.}
\label{tab:repository_experiment}
\begin{tabular}{lrrrrr}
\toprule
Repository &  N. PR & LOC & N. Test & Compile & Test Passes \\
\midrule
lottie-android & 2457 & 362 & 1 & Yes & No \\
mockito        & 3509 & 764 & 1 & Yes & Yes\\
aeron          & 1734 & 823 & 1 & No  & No\\
junit4         & 1765 & 198 & 1 & Yes & No\\
karatelabs     & 2415 & 493 & 2 & No  & No \\
grobid         & 1075 & 623 & 1 & Yes & Yes\\
jadx           & 2326 & 353 & 2 & No  & No\\
pinpoint       & 12591 & 631 & 1 & Yes & Yes\\
strimzi-kafka  & 11429 & 526 & 1 & Yes & Yes\\
vavr           & 3045  & 975 & 1 & No & No\\
\bottomrule
\end{tabular}
\end{table}

These results highlight two important aspects. First, {\approach} can be effectively used to instantiate realistic repository-level code generation experiments involving pull-request interpretation, repository navigation, multi-file modification, compilation, testing, and qualitative inspection. Second, the results reported in Table~\ref{tab:repository_experiment} show a gap between the implementation of the requested modification and successful test execution, suggesting that the selected pull requests remain challenging for current agent-based code generation systems.

\section{Conclusions} \label{sec:conclusions}
AI-based code generation is gradually evolving toward increasingly automated tools capable of addressing code generation tasks at progressively larger scales. This paper presents {\approach}, a curated dataset of pull requests and associated data that provides a foundation for investigating key research questions in AI-based code generation and software evolution. As reported in the paper, {\approach} includes a large collection of heterogeneous cases, including large PRs spanning many files and locations, as well as modifications involving non-source code artifacts. The dataset is available publicly for use and experimentation~\cite{repository}. Future work will explore the use of {\approach} to identify the capabilities and limitations of state-of-the-art AI-based techniques, and to support the development of next-generation models capable of  handling realistic, feature-level software changes.

\section*{Acknowledgment}
This work was partially supported by the MUR under the grant “Dipartimenti di Eccellenza 2023-2027" of the Department of Informatics, Systems and Communication of the University of Milano-Bicocca, Italy; and by the Engineered MachinE Learning-intensive IoT systems (EMELIOT) national research project, which has been funded by the MUR under the PRIN 2020 program (Contract 2020W3A5FY).

\bibliographystyle{unsrt}
\bibliography{bib}

\begin{IEEEbiography}[{\includegraphics[width=1in,height=1.25in,clip,keepaspectratio]{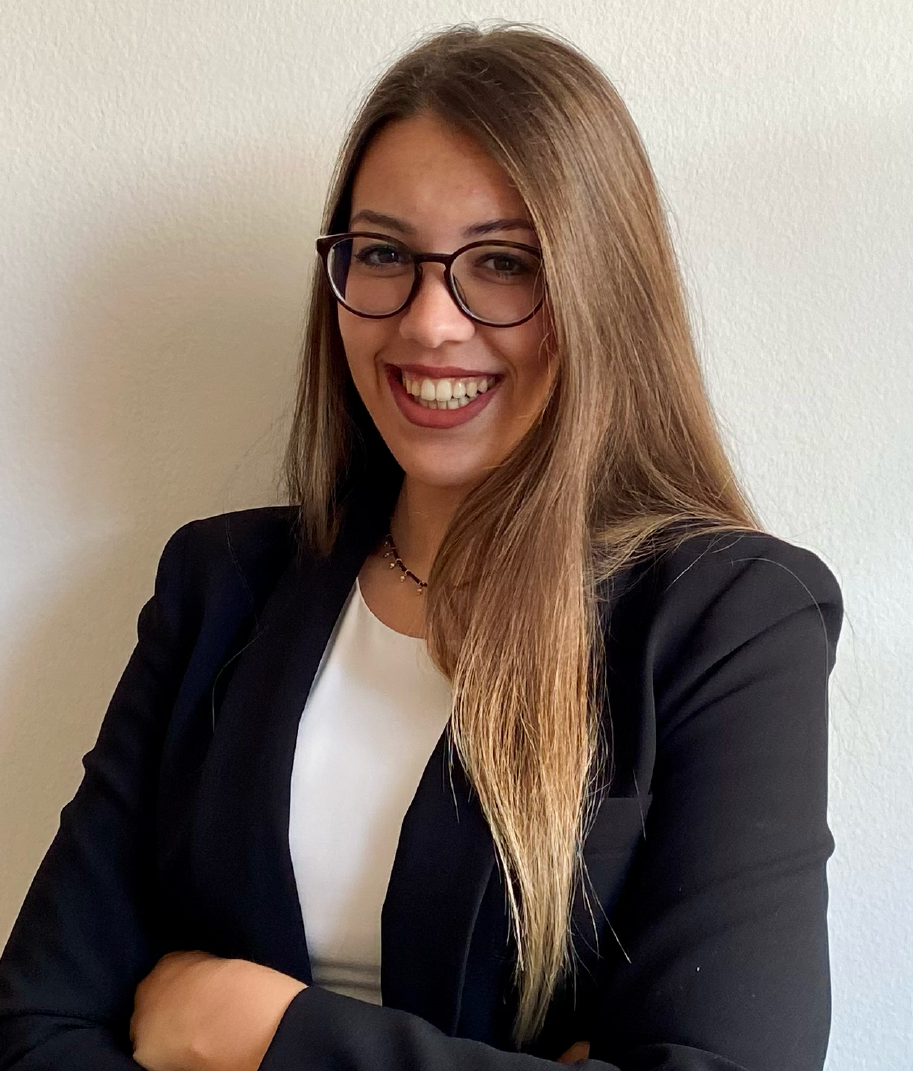}}]{Benedetta Donato} is a Ph.D. student in Computer Science at the University of Milano-Bicocca. Her current research interests include AI-assisted software engineering, Large Language Models for code generation, and empirical studies on developer–AI collaboration. 
She received her 
Master’s degree in Computer Science in 2024, with a thesis focused on the impact of generation parameters on the quality and correctness of code produced by LLMs, awarded by the University of Milano-Bicocca. She's an IEEE Student Member. Contact information: benedetta.donato@unimib.it. 
\end{IEEEbiography}

\begin{IEEEbiography}[{\includegraphics[width=1in,height=1.25in,clip,keepaspectratio]{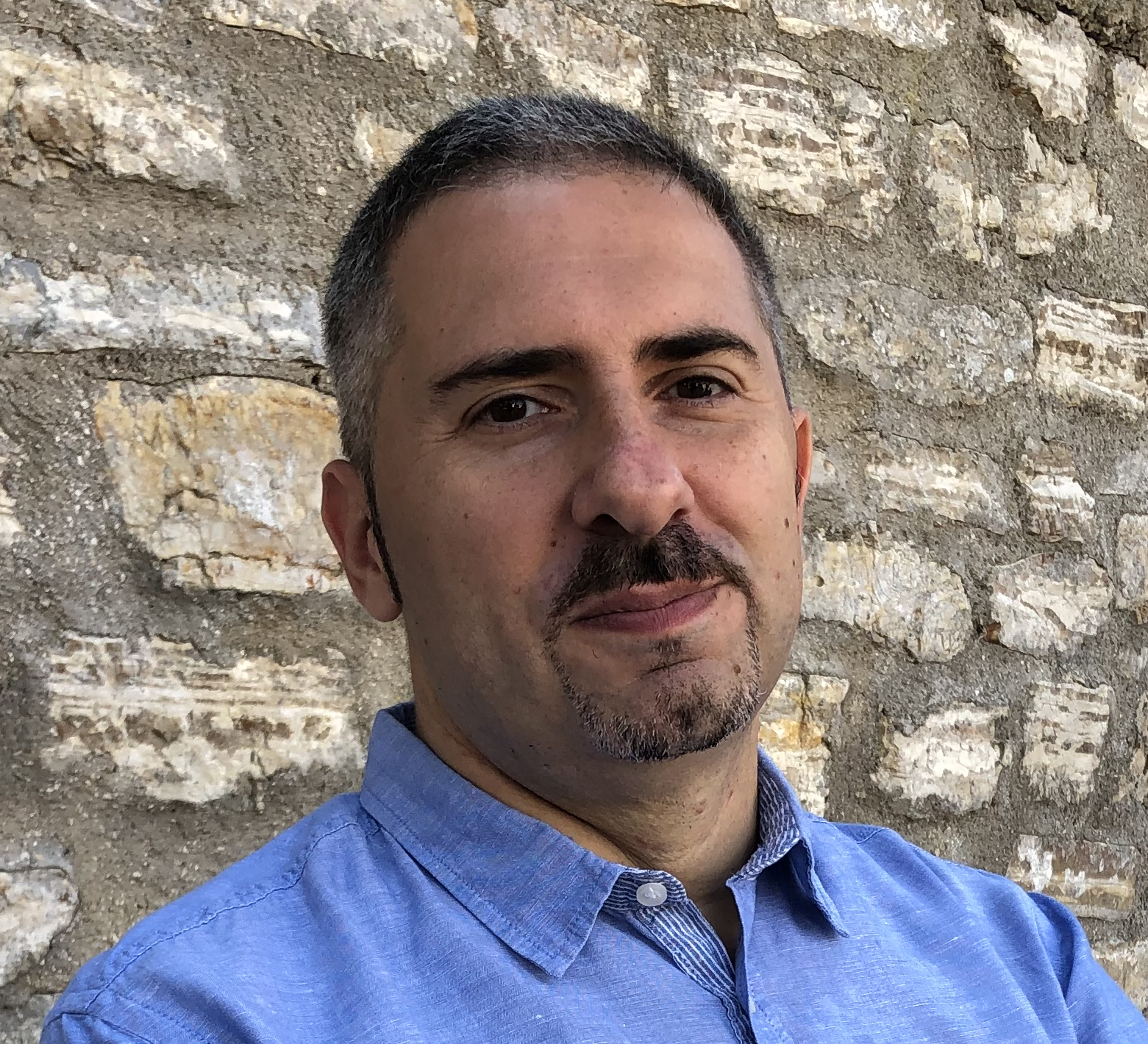}}]{Leonardo Mariani} is a Full Professor at the University of Milano-Bicocca. He holds a PhD in Computer Science, received from the same university in 2005. His research interests include software engineering, in particular software testing, program analysis, automated debugging, AI4SE, and self-adaptive systems. He has authored more than 150 papers appeared at top software engineering conferences and journals. He received the ICST MIP award for his ICST
2013 paper. He was awarded the ERC Consolidator Grant in 2015, an ERC Proof of Concept grant in 2018, and he is currently active in several national and international projects. He is regularly involved in the organization of major software engineering conferences.
\end{IEEEbiography}

\newpage

\begin{IEEEbiography}[{\includegraphics[width=1in,height=1.25in,clip,keepaspectratio]{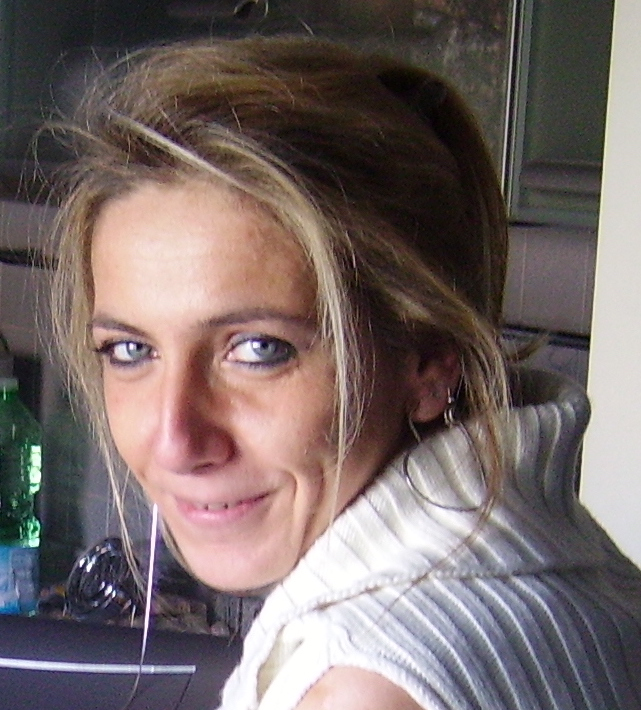}}]{Daniela Micucci} received the Ph.D. degree in Mathematics, Statistics, Computational Sciences and Computer Science (Informatics curriculum) from the University of Milano in 2004. She is currently Associate Professor with the Department of Informatics, Systems and Communication, University of Milano-Bicocca, where she leads the Software Architecture Laboratory (SAL). Her research interests include software architecture, software reliability, digital health systems, and the application of large language models to software engineering. She has participated and is currently participating in several national and European research projects and has authored numerous publications in leading international journals and conferences (e.g., IEEE TSE, ICSE, ASE, ISSRE). She serves as editor and reviewer for major venues in software engineering and intelligent environments. 
\end{IEEEbiography}

\begin{IEEEbiography}[{\includegraphics[width=1in,height=1.25in,clip,keepaspectratio]{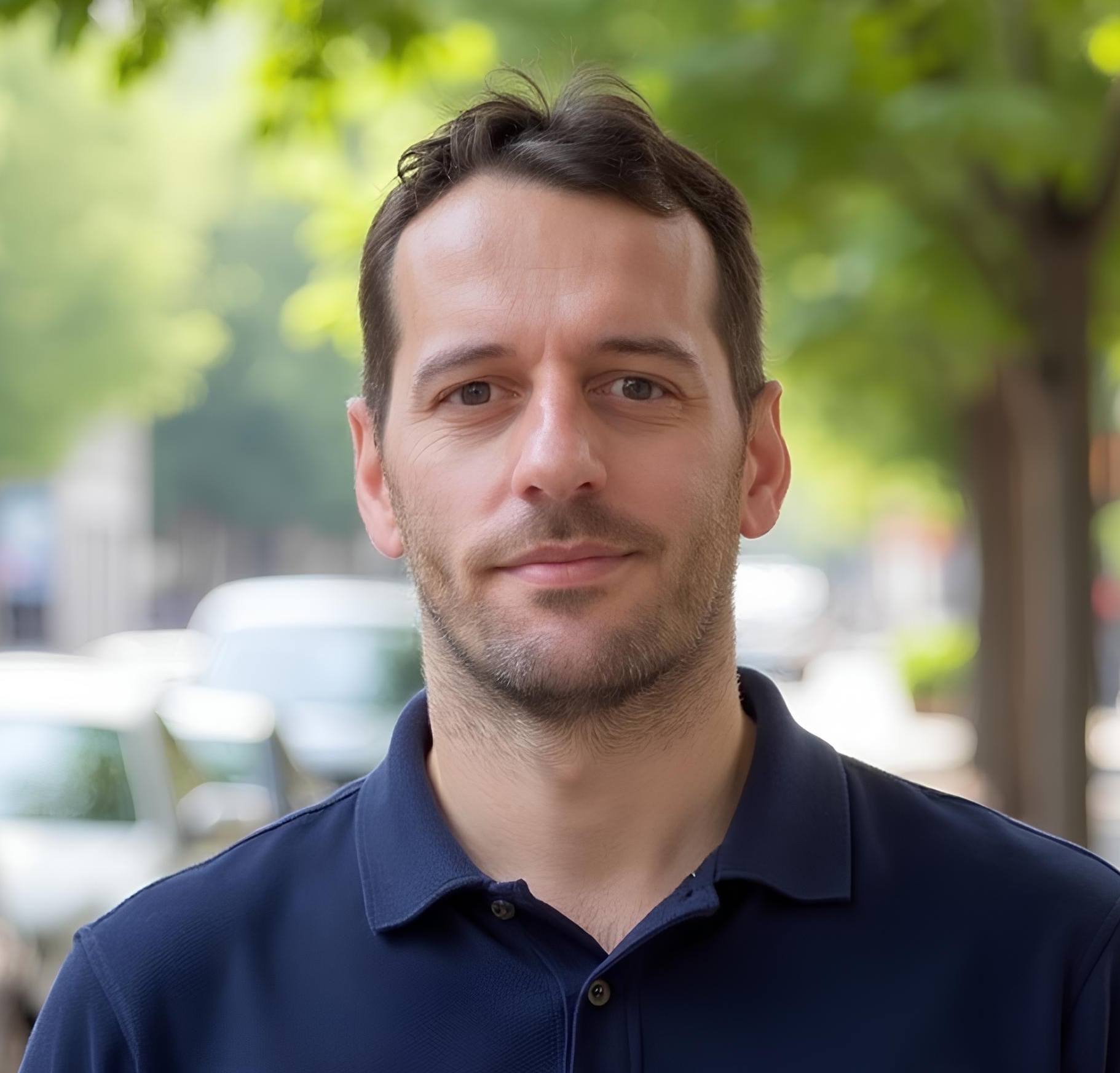}}]{Oliviero Riganelli} is an Associate Professor with the Department of Informatics, Systems and Communication at the University of Milano-Bicocca, Italy. He received the Ph.D.\ degree in Computer Science and Complex Systems from the University of Camerino, Italy. He has also conducted research at the University of Lugano (Switzerland) and Stony Brook University (United States). His research spans a wide range of topics in software engineering, with a focus on methodologies and technologies to improve the correctness, reliability, and quality of software systems. In particular, his work covers software testing, program analysis, AI-assisted software development, cloud computing, and autonomous systems. He has participated in several national and international research projects, collaborating with academic institutions and industrial partners. He is actively engaged in the software engineering research community, serving as a reviewer and program committee member for leading conferences and journals. His research contributions have been published in prominent international venues in software engineering. 
\end{IEEEbiography}

\EOD

\end{document}